\documentclass[journal,transmag]{IEEEtran}
\usepackage[margin=1in]{geometry}
\usepackage{color}
\usepackage{amssymb}
\usepackage{amsmath}
\usepackage{fixltx2e} 		
\usepackage{graphicx}
\usepackage{lineno} 			
\usepackage{siunitx}
\usepackage{amsthm}
\usepackage{epstopdf}
\usepackage{tocloft}
\usepackage{todonotes}

\renewcommand*\cftfigpresnum{Figure~}
\usepackage{float}

\usepackage[hidelinks]{hyperref}
\makeatletter
\providecommand{\doi}[1]{%
  \begingroup
    \let\bibinfo\@secondoftwo
    \urlstyle{rm}%
    \href{http://dx.doi.org/#1}{%
      doi:\discretionary{}{}{}%
      \nolinkurl{#1}%
    }%
  \endgroup
}
\makeatother

\ifCLASSINFOpdf
\else
\fi
\newtheorem{mydef}{Definition}
\newtheorem{mythe}{Theorem}

\begin{document}
%

\title{Observability analysis and state estimation of lithium-ion batteries in the presence of sensor biases}

\author{\IEEEauthorblockN{Shi Zhao,
Stephen R. Duncan, \emph{Member, IEEE},
and David A. Howey, \emph{Member, IEEE}}
\thanks{Manuscript received October 5, 2015; revised February 1, 2016; accepted March 5, 2016. This work is funded through the RCUK Energy Programme's STABLE-NET project under Grant EP/L014343/1. \emph{(Corresponding author: Shi Zhao.)}

The authors are with the Department~of~Engineering~Science, University~of~Oxford, Parks~Road, Oxford, OX1~3PJ, United~Kingdom.(email: shi.zhao@eng.ox.ac.uk; stephen.duncan@eng.ox.ac.uk; david.howey@eng.ox.ac.uk).}}

\IEEEtitleabstractindextext{%
\begin{abstract}
This brief investigates the observability of one of the most commonly used equivalent circuit models (ECMs) for lithium-ion batteries and presents a method to estimate the state of charge (SOC) in the presence of sensor biases, highlighting the importance of observability analysis for choosing appropriate state estimation algorithms. Using a differential geometric approach, necessary and sufficient conditions for the nonlinear ECM to be observable are derived and are shown to be different from the conditions for the observability of the linearised model. It is then demonstrated that biases in the measurements, due to sensor ageing or calibration errors, can be estimated by applying a nonlinear Kalman filter to an augmented model where the biases are incorporated into the state vector. Experiments are carried out on a lithium-ion pouch cell and three types of nonlinear filters, the first-order extended Kalman filter (EKF), the second-order EKF and the unscented Kalman filter (UKF) are applied using experimental data. The different performances of the filters are explained from the point of view of observability.
\end{abstract}

\begin{IEEEkeywords}
battery, equivalent circuit model, Kalman filtering, observability, sensor bias, state estimation.
\end{IEEEkeywords}}


\markboth{IEEE Transactions on Control Systems Technology}%
{Shell \MakeLowercase{\textit{et al.}}: Bare Demo of IEEEtran.cls for Journals}
%





\maketitle

\IEEEdisplaynontitleabstractindextext

%
\IEEEpeerreviewmaketitle

\section{Introduction}
A battery management system (BMS) consists of hardware and software that ensure the safe charging and discharging of battery cells \cite{Howey2015}. A key function of a BMS is accurate estimation of the state of charge (SOC). As battery SOC is not directly measureable, it has to be inferred from the available measurements, such as the applied current, terminal voltage and surface temperature. A mathematical model that characterises the dynamics of a battery is essential for state estimation.

Generally speaking, there are two types of battery models: equivalent circuit models (ECMs) and electrochemical models. ECMs are low order models parameterised from time domain or frequency domain experimental data using system identification techniques. The accuracy of an ECM is usually quantified by how closely the model output matches the experimental measurements when the battery is under a dynamic current load \cite{hu2012comparative}. On the other hand, electrochemical models such as the pseudo two-dimensional (P2D) model \cite{doyle1993modeling}, are derived from electrochemical principles and usually characterised by partial differential equations coupled with algebraic constraints (PDAEs) that describe the thermodynamics, reaction kinetics and transport within a battery cell.

Due to the high computational complexity of applying state and parameter estimation algorithms to PDAEs, electrochemical models are more difficult to implement in a BMS,
despite the fact that they are able to accurately describe the battery dynamics in a wider operating range. 
There have been some attempts to estimate the battery states using simplified electrochemical models. For example, the single particle model (SPM), a simplification of the P2D model, is used in many studies for state estimation \cite{Moura2013b,DiDomenico2010,Bartlett2015a}. The problem with the SPM is that it is only valid under low C-rates and the accuracy deteriorates when the applied current becomes highly dynamic.
State estimation of the full P2D model using a modified extended Kalman filter (EKF) has recently been reported in \cite{Bizeray2015lithium}, where the electrochemical model is solved by the Chebyshev orthogonal collocation method to reduce the computational cost. However, parameter estimation of the model remains a significant challenge.

For these reasons, currently a typical BMS uses an ECM, although the parameters of the model are less easy to interpret in relation to the physics. SOC estimation based on ECMs has been studied extensively and  one major theme for research in this area is to improve the accuracy of an ECM by either employing more sophisticated parametrisation algorithms or increasing the complexity of the model itself  \cite{hu2012comparative,plett2004a}. As for state estimation algorithms, nonlinear Kalman filters such as the first-order EKF and sigma point Kalman filters are among the most popular choices \cite{plett2004a, plett2006a}.

Although there are some articles that examine the observability of battery models \cite{Moura2013b} \cite{lin2015state}, generally, it is an overlooked topic. In many papers on battery state estimation, the observability of the models is not considered \cite{hu2012comparative}. Observability analysis is important because it is not possible to estimate the states of a battery if the model is not observable, regardless how well the model matches the input-output behaviour of the battery. For this reason, when choosing an ECM for a BMS, the emphasis should not be placed solely on finding the model that gives smallest output error compared to experimental measurements; the observability of the model should also be considered.
Moreover, even if the model is observable, the state estimation algorithm still needs to be chosen with care to account for the reason for the model observability. For example, if the nonlinear model is locally observable everywhere but its linearised version is not observable, then a first-order EKF may fail to track the SOC accurately \cite{lin2015state}.

In this brief, we use a differential geometric approach to
analyse the nonlinear observability of a second-order RC model for batteries and derive the necessary and sufficient condition for the model to be observable. It is shown that the local observability of the model at a particular state is different from the observability of the linearised model around this state. We then show that the battery states can still be estimated accurately even when there exist unknown biases in the sensors. This is because the augmented model which incorporates the unknown biases into the states is locally observable. Experiments are carried out on a lithium nickel manganese cobalt oxide (NMC) pouch cell (Kokam 740 mAh) and three types of nonlinear Kalman filters: the first-order EKF, the second-order EKF and the unscented Kalman filter (UKF) are applied to estimate the SOC of the cell using experimental data. The result shows that while the first-order EKF may not be able to 
estimate the SOC when there is a sensor bias, the estimation given by the UKF has a much better accuracy. The difference is explained by an observability analysis of the augmented model.

The brief is organised as follows. Section \ref{sec:obs_def} introduces the definition and criteria for nonlinear observability. The ECM under consideration is briefly discussed and its observability is analysed in Section \ref{sec:obs_ecm}. Section \ref{sec:obs_bias} presents observability analysis of augmented battery models with sensor biases incorporated into the states. Experimental validation is carried out in Section \ref{sec:val} and Section \ref{sec:conclusion} concludes.

\section{Nonlinear observability} \label{sec:obs_def}
The observability of a linear time-invariant system described by a state space model can be readily determined by checking whether the observability matrix has full rank. By contrast, the observability of a nonlinear system is less straightforward to determine. Although it is tempting to study the observability by linearising the system model, 
there are important differences between linear observability and nonlinear observability \cite{vidyasagar2002nonlinear}.

\subsection{Distinguishability and observability}
Denote $X$ as an open subset of $\mathbb{R}^n$. The nonlinear system under consideration is described by the following state space model
\begin{subequations} \label{sys:01}
\begin{eqnarray}
  \dot{x} &=& f(x) + \sum_{i=1}^{m} u_i g_i(x) \\
  y &=& h(x)
\end{eqnarray}
\end{subequations}
where $x \in X$ is the state, $u_i \in \mathbb{R}$ is the input, $y \in \mathbb{R}^p$ is the output, $f: X \rightarrow \mathbb{R}^n ,g_i: X \rightarrow \mathbb{R}^n$ and $h: X \rightarrow \mathbb{R}^p$ are all smooth functions. We now introduce the following definition \cite{vidyasagar2002nonlinear}. 

\begin{mydef}
Consider system (\ref{sys:01}) and two states $x_1$ and $x_2$. Denote the system output at time $t$ with initial state $x_i$ and input $u$  as $y(x_i,u,t)$ where $i=1,2$.
$x_1$ and $x_2$ are said to be distinguishable if there exists an input function $u$ such that $y(x_1,u,t) \neq y(x_2,u,t)$ for a finite $t$. System (\ref{sys:01}) is locally observable at $x_1$ if there exists a neighbourhood $\mathcal{N}$ of $x_1$ such that the only state in $\mathcal{N}$ that is not distinguishable from $x_1$ is $x_1$ itself. The system is said to be locally observable if it is observable at every $x \in X$.
\end{mydef}
One subtlety is that two states of a nonlinear system may be distinguishable even if $y(x_1,u,t) = y(x_2,u,t)$ for some input functions $u$. By contrast, for a linear system, $y(x_1,u,t) \neq y(x_2,u,t)$ should hold for any $u$ if the two states are distinguishable. This means that in general it is more difficult to distinguish two states of a nonlinear system because it may not be trivial to find an input $u$ that gives different output functions. Moreover, for a linear system, local observability implies global observability. However, this is no longer the case for a nonlinear system.

\subsection{Observability rank test} \label{sec:2.2}
The observability of a nonlinear system can be determined by a rank test which involves Lie derivatives.
Suppose that $h(x)= \begin{bmatrix} h_1(x) & \ldots & h_p(x)  \end{bmatrix}^T$ is a $p$-dimensional vector function on $X$ and its $j$th component $h_j(x)$ is a real-valued smooth function. Denote the gradient of $h_j$ as ${\mathbf d} h_j$, i.e.,
\begin{equation*}
{\mathbf d} h_j =  \begin{bmatrix} \frac{\partial h_j}{\partial x_1} &\frac{\partial h_j}{\partial x_2} & \ldots & \frac{\partial h_j}{\partial x_n} \end{bmatrix}
\end{equation*}
then the Lie derivative of $h_j$ with respect to $f$ is a real-valued function defined by
\begin{equation*}
  L_f h_j =  {\mathbf d} h_j \cdot f = \sum_{i=1}^n f_i \frac{\partial h_j}{\partial x_i}
\end{equation*}
where $f(x)= \begin{bmatrix} f_1(x) & \ldots & f_n(x)  \end{bmatrix}^T$ . The zeroth order Lie derivative $L_f^0 h_j$ is $h_j$ itself and the second-order Lie derivative $L_f^2 h_j$ is defined by $L_f^2 h_j = L_f L_f h_j$.

The following theorem \cite{vidyasagar2002nonlinear} gives the rank test for local observability of the nonlinear system.
\begin{mythe} \label{the:01}
System (\ref{sys:01}) is locally observable at $x_0 \in X$ if there are $n$ linearly independent rows in the set
\begin{equation*}
({\mathbf d} L_{z_s} L_{z_{s-1}}\ldots L_{z_1} h_j) (x_0)
\end{equation*}
where $s \geq 0$, $z_k \in \{f, g_1,\ldots, g_m\}$ for $k=1,\ldots, s$, $j=1,\ldots, p$ and with $s=0$, the expression is defined as equivalent to ${\mathbf d} h_j(x_0)$.
\end{mythe}

The observability rank test for a linear system can be derived from this theorem. It should be noted that in the theorem there is no upper bound for $s$, which means that the number of Lie derivatives in the calculation is not bounded from above for a general nonlinear system (\ref{sys:01}). In addition, the rank condition given by the theorem is only a sufficient but not a necessary condition for system (\ref{sys:01}) to be locally observable.

\section{Observability of an ECM} \label{sec:obs_ecm}
We now study the observability of battery models using the rank test introduced in the previous section.
\subsection{Battery ECM}
\begin{figure}
\centering
\includegraphics[width=0.4\textwidth]{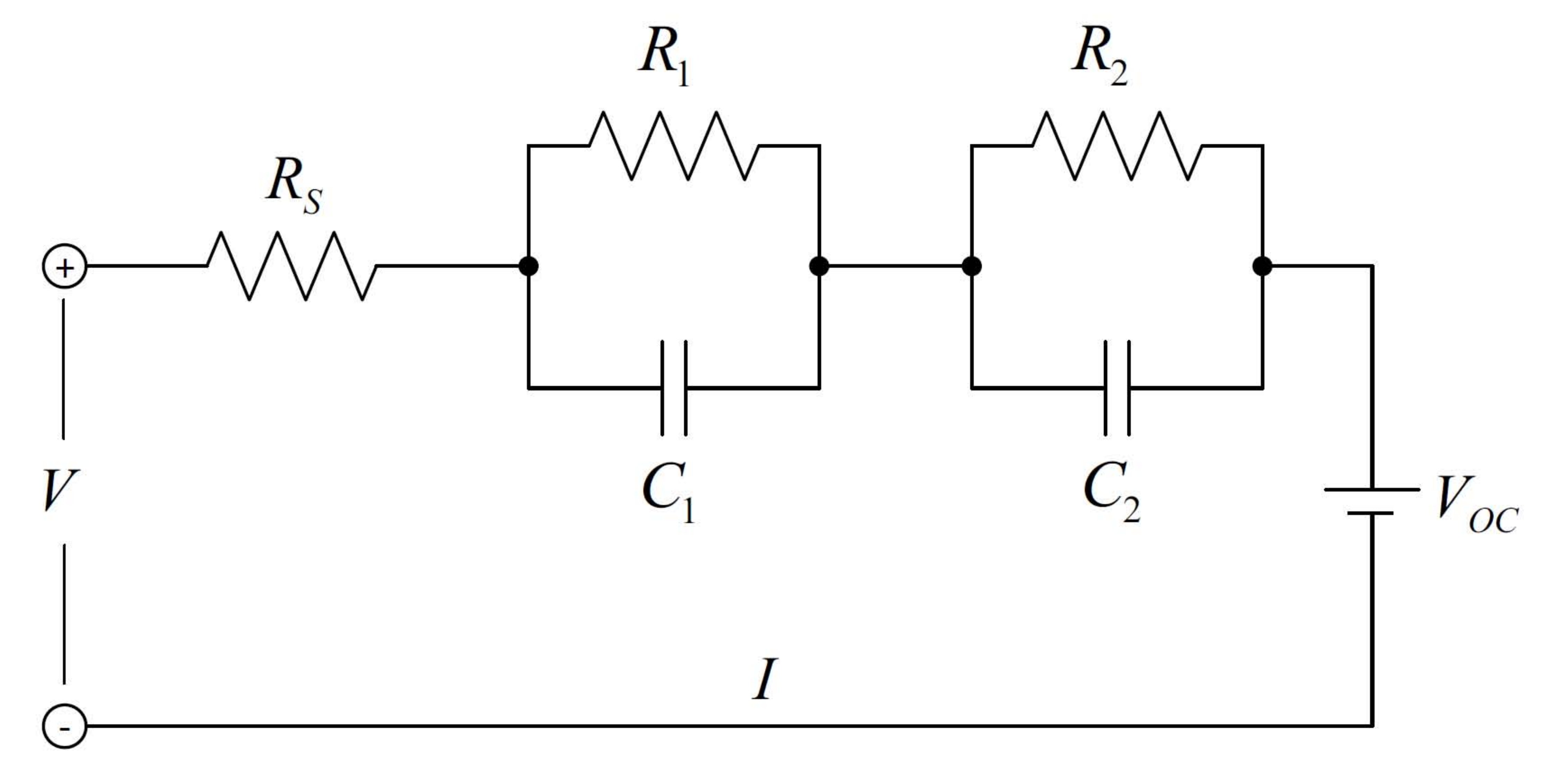}
\caption{Schematic diagram of the ECM.}
\label{fig:Figure1}
\end{figure}

The model considered here is the second-order RC model shown in Fig. \ref{fig:Figure1}, which is one of the most widely used ECMs in the literature \cite{hu2012comparative}. The open circuit voltage (OCV) $V_{OC}$ is a nonlinear function of the SOC and Weng \emph{et. al} \cite{weng2014unified} summarise some of the popular OCV models, which are all smooth nonlinear functions. The $V_{OC}$ is fitted to approximate the OCV-SOC data collected by using the galvanostatic intermittent titration technique (GITT) or charging and discharging a battery cell at a low constant current (e.g. 1/25C) under constant temperature. The parameters $R_1,R_2,C_1,C_2$ and $R_s$ are identified such that the model output matches the experimental measurements as closely as possible.
In this model, these parameters are set to be constants, but we note that the model accuracy can be improved if the parameters are updated iteratively and allowed to vary with the SOC and temperature, at the cost of higher computational complexity. If the parameters are state-dependent, the observability of the model can still be analysed using Theorem \ref{the:01}, although the computation becomes more laborious.

The state space representation of the model in the continuous form is given by
\begin{subequations}\label{sys:02}
\begin{eqnarray}
  \begin{bmatrix} \dot{V}_1 \\ \dot{V}_2 \\ \dot{Z} \end{bmatrix} &=&  \begin{bmatrix} -\frac{1}{R_1 C_1} & 0 & 0 \\ 0 & -\frac{1}{R_2 C_2} & 0 \\ 0 & 0 & 0 \end{bmatrix} \begin{bmatrix} V_1 \\ V_2 \\ Z \end{bmatrix} + \begin{bmatrix} \frac{1}{C_1} \\ \frac{1}{C_2} \\ -\frac{1}{Q}\end{bmatrix} I  \label{sys:02a} \nonumber \\
  \\
  V &=& V_{OC}(Z) - V_1 - V_2 - I R_s \label{sys:02b}
\end{eqnarray}
\end{subequations}
where $V_1,V_2$ are the voltages across the first and the second RC pairs, respectively, $Z \in [\SI{0}{\percent},\SI{100}{\percent}]$ is the normalised SOC, $Q$ is the battery capacity, $I$ is the current applied to the battery and $V$ is the terminal voltage. Note that the sign of the input $I$ is taken to be positive when the battery is discharging and negative during charging.
For ease of reference, define $\tau_1 = R_1 C_1$ and $\tau_2 = R_2 C_2$.

The state equation (\ref{sys:02a}) is linear and the only nonlinearity in the model comes from the $V_{OC}$ in the output equation (\ref{sys:02b}). To apply the nonlinear observability theorem to the system, we first rewrite the model in the following form
\begin{subequations} \label{form:02}
\begin{eqnarray}
  \dot{x} &=& f(x) + g u  \\
  y &=& h(x) -  R_s u 
\end{eqnarray}
\end{subequations}
where $x = [V_1, V_2, Z]^T$,  $u = I, y = V,  f(x) = \begin{bmatrix} -\frac{1}{\tau_1} V_1 &  -\frac{1}{\tau_2}V_2 & 0\end{bmatrix}^T$,  $g = \begin{bmatrix}\frac{1}{C_1} & \frac{1}{C_2} & -\frac{1}{Q} \end{bmatrix}^T$ and $h(x) = V_{OC}(Z)-V_1-V_2$. The feedthrough term $R_s u$ in the output equation does not affect the applicability of Theorem \ref{the:01} to the model.

\subsection{Observability analysis} \label{sec:ecm_obs}
Using the notation introduced in Section \ref{sec:2.2}, the gradient of $h(x)$ with respect to $x$ is
\begin{equation*}
{\mathbf d} h = \begin{bmatrix} -1 & -1 & \frac{{\rm d} V_{OC}}{{\rm d} Z} \end{bmatrix}
\end{equation*}
And it can be shown by mathematical induction that
\begin{eqnarray*}
 {\mathbf d} L_f^{k} h &=& \begin{bmatrix} -\frac{1}{(-\tau_1)^{k}} & -\frac{1}{(-\tau_2)^{k}} & 0  \end{bmatrix} \\
 {\mathbf d} L_g^{k} h &=& \begin{bmatrix} 0 & 0 & \frac{1}{(-Q)^k}  \frac{{\rm d}^{k+1} V_{OC}}{{\rm d}Z^{k+1}} \end{bmatrix}
\end{eqnarray*}
for all $k \in \mathbb{Z}^{+}$ and Lie derivatives involving both $f$ and $g$ such as $L_{f}L_{g}h$ are constants, thus can be ignored in the rank test.

Model (\ref{sys:02}) is locally observable at a point $x_0$ if the set
\begin{equation}\label{set:o}
\mathcal{O} = \begin{bmatrix}{\mathbf d} h \\  {\mathbf d} L_f h \\ {\mathbf d} L_g h  \\ {\mathbf d} L_f^2 h \\  {\mathbf d} L_g^2 h \\ \vdots  \end{bmatrix}
\end{equation}
evaluated at $x_0$ has $n=3$ linearly independent row vectors. One thing to notice is that the rank condition does not hold if $\tau_1 = \tau_2$. Physically, this is because when $\tau_1 = \tau_2$, the two RC pairs can be combined into a single one, so that the voltage across either RC pair cannot be uniquely determined. In practice, the two time constants are of different magnitudes and in the following analysis we assume that $\tau_1 \neq \tau_2$.

Under this assumption, $\mathcal{O} (x_0)$ has full column rank if and only if there exists a $k \in \mathbb{Z}^{+}$ such that
\begin{equation} \label{con:01}
  \left(\frac{{\rm d}^{k} V_{OC}}{{\rm d}Z^{k}} \right) (x_0) \neq 0
\end{equation}
This means that the battery model (\ref{sys:02}) is observable if all the derivatives of $V_{OC}$ 
 are not zero simultaneously.

As an aside, note that since the state equation (\ref{sys:02a}) is linear and the system is controllable, according to Proposition 3.38 in \cite{nijmeijer1990nonlinear}, the rank condition is also a necessary condition for model (\ref{sys:02}) to be locally observable at $x_0$. Thus if (\ref{con:01}) is not satisfied, the system is not locally observable.

In the literature, it is common to simply equate the local observability of a nonlinear system with the observability of its linearised system. Linearising model (\ref{sys:02}) at $x_0$ gives
\begin{eqnarray*}
  \dot{x} &=& A x + Bu \\
  y &=& Cx + Du
\end{eqnarray*}
where
\begin{eqnarray*}
A = \begin{bmatrix} -\frac{1}{\tau_1} & 0 & 0 \\ 0 & -\frac{1}{\tau_2} & 0 \\ 0 & 0 & 0 \end{bmatrix} ,~~ B = \begin{bmatrix} \frac{1}{C_1} \\ \frac{1}{C_2} \\ \frac{1}{-Q} \end{bmatrix} \\ C = \begin{bmatrix} -1 & -1 & \left(\frac{{\rm d}V_{OC}}{{\rm d}Z}\right)(x_0) \end{bmatrix},~~ D = R_s
\end{eqnarray*}
The observability of this linearised model can be determined by checking the rank of the observability matrix
\begin{equation*}
 \begin{bmatrix} C \\ CA \\ CA^2 \end{bmatrix} = \begin{bmatrix} -1 & -1 & \left(\frac{{\rm d}V_{OC}}{{\rm d}Z}\right)(x_0)  \\  \frac{1}{\tau_1} & \frac{1}{\tau_2} & 0 \\ -\frac{1}{\tau_1^2} & -\frac{1}{\tau_2^2} & 0 \end{bmatrix}
\end{equation*}
This matrix has full rank if and only if $\left(\frac{{\rm d}V_{OC}}{{\rm d}Z}\right)(x_0) \neq 0$. However, this is not a necessary condition for nonlinear system (\ref{sys:02}) to be locally observable at $x_0$, so that system (\ref{sys:02}) may be observable even if its linearisation is not.

\section{Estimation with sensor biases} \label{sec:obs_bias}

The accuracy of state estimation is known to be dependent on the quality (precision and accuracy) of the sensors that give the input and output measurements. It is commonly assumed that the current and voltage sensors in model (\ref{sys:02}) are subject to zero mean Gaussian noise, where the sensor precisions are described by the variances of the Gaussian distributions. However, the sensors may also have some unknown biases due to ageing or calibration errors, which is referred to as a sensor bias fault and there are mainly two ways to detect the fault: hardware redundancy and analytical redundancy \cite{vemuri2001sensor}. In the hardware redundancy approach, redundant sensors are used to measure one variable and a fault is detected if there is notable difference between the measurements from different sensors. In the analytical redundancy approach, the system is analysed using multiple model based techniques.

When there is a sensor bias fault, it is expected that the accuracy of the state estimation deteriorates when the effect of a sensor bias builds up over time. One approach that is able to estimate the sensor biases and the battery SOC at the same time, without incorporating any additional sensors into the system, is to augment the model \cite{gustafsson2000adaptive}. As in \cite{vemuri2001sensor, gustafsson2000adaptive}, the sensor biases are modelled as constants here.

\subsection{ECM with sensor biases}
We first consider the case that there is an unknown bias in the voltage sensor, while there is no bias in the current sensor. Denote the voltage sensor bias as $V_e$ and the inaccurate voltage measurement as $V_m$, then $V_m-V_e$ represents the actual battery terminal voltage denoted as $V$. The difficulty is the bias $V_e$ is unknown and if we simply ignore it, the estimation based on model (\ref{sys:02}) and $V_m$ will be inaccurate.
The problem can be solved by including the bias $V_e$ into the state vector and reformulating the model. Specifically, since $V = V_m - V_e$,
model (\ref{sys:02}) can now  be reformulated as
\begin{subequations} \label{sys:2RC_output}
\begin{equation}
 \begin{bmatrix} \dot{V}_1 \\ \dot{V}_2 \\ \dot{Z} \\ \dot{V}_e \end{bmatrix} =  \begin{bmatrix} -\frac{1}{\tau_1} & 0 & 0 & 0 \\ 0 & -\frac{1}{\tau_2} & 0 & 0 \\ 0 & 0 & 0 & 0 \\ 0& 0 & 0 & 0 \end{bmatrix} \begin{bmatrix} V_1 \\ V_2 \\ Z \\V_e \end{bmatrix} + \begin{bmatrix} \frac{1}{C_1} \\ \frac{1}{C_2} \\ -\frac{1}{Q} \\ 0\end{bmatrix} I
 \end{equation}
 \begin{equation}
  V_{m} = V_{OC}(Z) - V_1 - V_2 - I R_s + V_e
\end{equation}
\end{subequations}
Such a model is called an augmented model, while (\ref{sys:02}) is called the original model.

The augmented model can be rewritten into the form in (\ref{form:02}) where $x = [V_1, V_2, Z, V_e]^T$,  $u = I, y = V_m,  f(x) = \begin{bmatrix} -\frac{1}{\tau_1} V_1 &  -\frac{1}{\tau_2}V_2 & 0 & 0\end{bmatrix}^T$,  $g = \begin{bmatrix}\frac{1}{C_1} & \frac{1}{C_2} & -\frac{1}{Q} & 0 \end{bmatrix}^T$ and $h(x) = V_{OC}(Z)-V_1-V_2 + V_e$. Note that the augmented model is not controllable.

For the case that there is an unknown bias in the current sensor but no bias in the voltage sensor, the model can be reformulated in the same manner. Denote the bias as $I_e$ and the inaccurate current measurement as $I_m$, then $I = I_m-I_e$ is the actual applied current. As a result, the battery model (\ref{sys:02}) now becomes
\begin{subequations} \label{sys:2RC_input}
\begin{equation}
\begin{bmatrix} \dot{V}_1 \\ \dot{V}_2 \\ \dot{Z} \\ \dot{I}_e \end{bmatrix} =  \begin{bmatrix} -\frac{1}{\tau_1} & 0 & 0 & -\frac{1}{C_1} \\ 0 & -\frac{1}{\tau_2} & 0 & -\frac{1}{C_2}\\ 0 & 0 & 0 & \frac{1}{Q} \\ 0& 0 & 0 & 0 \end{bmatrix} \begin{bmatrix} V_1 \\ V_2 \\ Z \\I_e \end{bmatrix} + \begin{bmatrix} \frac{1}{C_1} \\ \frac{1}{C_2} \\ -\frac{1}{Q} \\ 0\end{bmatrix} I_m
\end{equation}
\begin{equation}
  V = V_{OC}(Z) - V_1 - V_2 - I_m R_s + I_e R_s
\end{equation}
\end{subequations}

When there are biases in both the current and voltage sensors, an augmented model including both $I_e$ and $V_e$ in the state vector can be easily derived. For brevity, the model is not given here. In the following analysis, the focus is mainly on the first case, which is described by model (\ref{sys:2RC_output}). The analysis can be readily extended to other cases and is not repeated.

\subsection{Observability analysis} \label{sec:4.2}
It can be shown that for model (\ref{sys:2RC_output})
\begin{eqnarray*}
  {\mathbf d} h &=& \begin{bmatrix} -1 & -1 & \frac{{\rm d} V_{OC}}{{\rm d} Z} & 1 \end{bmatrix} \\
  {\mathbf d} L_f^k h &=& \begin{bmatrix} -\frac{1}{(-\tau_1)^{k}} & -\frac{1}{(-\tau_2)^{k}} & 0 & 0 \end{bmatrix} \\
  {\mathbf d} L_g^k h &=& \begin{bmatrix} 0& 0 & \frac{1}{(-Q)^{k}} \frac{{\rm d}^{k+1} V_{OC}}{{\rm d} Z^{k+1}} & 0\end{bmatrix}
\end{eqnarray*}
for all $k \in \mathbb{Z}^{+}$. Once again, Lie derivatives involving both $f$ and $g$ can be ignored in the rank test since they are all constants.

Model (\ref{sys:2RC_output}) is locally observable at a point $x_0$ if the set $\mathcal{O}$ (see (\ref{set:o}) in \ref{sec:ecm_obs}) evaluated at $x_0$ has $n=4$ linearly independent row vectors. This rank condition is satisfied if and only if there exists an integer $k \geq 2$ such that
\begin{equation} \label{con:02}
  \left(\frac{{\rm d}^{k} V_{OC}}{{\rm d}Z^{k}} \right) (x_0) \neq 0
\end{equation}

If condition (\ref{con:02}) does not hold, then ${\mathbf d} L_g^i$ is a zero row vector for any $i \in \mathbb{Z}^{+}$, which means that $\mathcal{O}(x_0)$ can at most have rank $3$. Therefore the necessity is proved. Now suppose the condition is satisfied, the following matrix
\begin{equation*}
  \begin{bmatrix} {\mathbf d} h \\ {\mathbf d} L_f h \\ {\mathbf d} L_f^2 h \\ {\mathbf d} L_g^{k-1} h \end{bmatrix} = \begin{bmatrix} -1 & -1 & \frac{{\rm d} V_{OC}}{{\rm d} Z}(x_0) & 1 \\ \frac{1}{\tau_1} & \frac{1}{\tau_2} & 0 & 0 \\ -\frac{1}{\tau_1^{2}} & -\frac{1}{\tau_2^{2}} & 0 & 0 \\  0& 0 & \frac{1}{(-Q)^{k-1}} \frac{{\rm d}^{k} V_{OC}}{{\rm d} Z^{k}}(x_0)& 0 \end{bmatrix}
\end{equation*}
has full rank. This can be verified by calculating the determinant of the matrix. Thus the condition is also a sufficient one.

We note that since system (\ref{sys:2RC_output}) is not controllable, the rank condition is no longer a necessary condition for the system to be locally observable at $x_0$. Therefore condition (\ref{con:02}) is a necessary and sufficient condition for $\mathcal{O}(x_0)$ to have full rank, which is only a sufficient condition for system (\ref{sys:2RC_output}) to be locally observable at $x_0$.

It is interesting to note that the observability of system (\ref{sys:2RC_output}) crucially depends on the nonlinearity of $V_{OC}$. If the OCV is a linear function of SOC such as in a capacitor, the rank condition is not satisfied at any $x_0$ and system (\ref{sys:2RC_output}) is unobservable. 

Similar analysis can be applied to system (\ref{sys:2RC_input}). Although the calculation is more laborious, it turns out that the condition for system (\ref{sys:2RC_input}) to be locally observable at a point $x_0$ is the same as that for system (\ref{sys:2RC_output}).


\section{Experimental validation} \label{sec:val}
In this section, using experimental data, several nonlinear Kalman filters based on model (\ref{sys:2RC_output}) are implemented to estimate the states of a lithium-ion battery when there is an unknown bias in the voltage sensor.

\subsection{Battery test}
\begin{figure}
\centering
\includegraphics[width=0.46\textwidth]{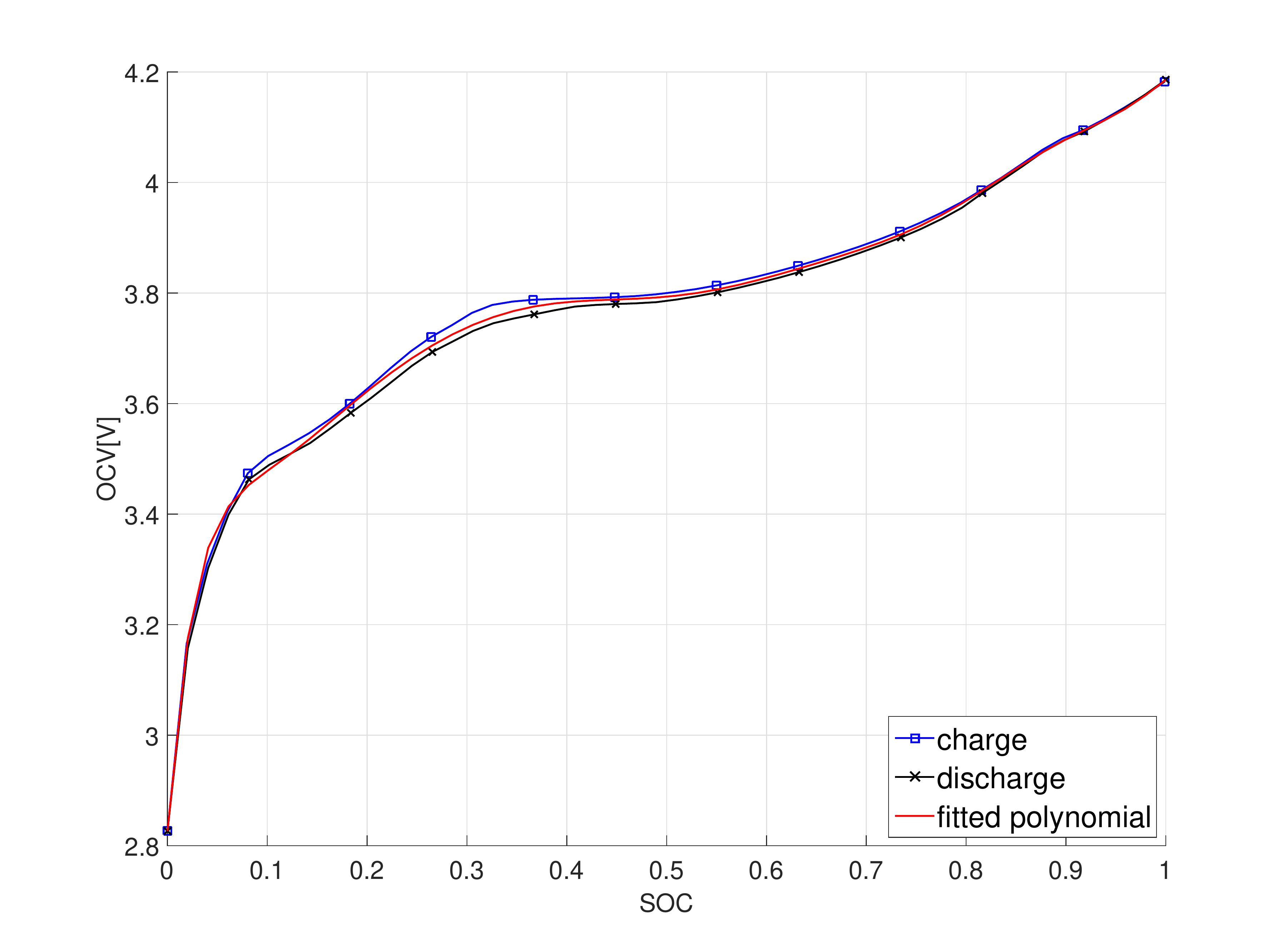}
\caption{OCV as a function of SOC.}
\label{fig:Figure2}
\end{figure}

The experiments were carried out on a 740 mAh Kokam NMC lithium-ion pouch cell (SLPB533459). The data were collected using a BioLogic SP-150 potentiostat and the temperature is fixed at $\SI{20}{\celsius}$ using a thermal chamber. A GITT procedure with current 0.1C was used to determine the relationship between the OCV and SOC, and 50 data points were recorded in both the charge and discharge processes \cite{birkl2015parametric}. As shown in Fig. \ref{fig:Figure2}, there is a discrepancy between the charge and discharge curves due to hysteresis.
The average of these two curves is taken as the OCV-SOC function and is approximated by a polynomial $V_{OC} = \sum_{k=0}^{12} a_k Z^{k}$ where the coefficients are given in Table \ref{tab:01} and the fitted polynomial is plotted in Fig. \ref{fig:Figure2}. The maximal approximation error of this polynomial in the SOC range $[10\%,100\%]$ compared to the averaged OCV-SOC curve is $\SI{17.97} {\milli\volt}$ and the root-mean-square error (RMSE) is $\SI{3.85} {\milli\volt}$.

\begin{table}
\centering
\caption{The values of the OCV polynomial coefficients.}
\begin{tabular}{|c|c|}
\hline
 coefficient & value \\
  \hline
  $a_0$ & $2.83$ \\
  $a_1$ & $2.41 \times 10^1$ \\
  $a_2$ & $-4.19 \times 10^2$ \\
  $a_3$ & $4.28 \times 10^3$ \\
  $a_4$ & $-2.73 \times 10^4$\\
  $a_5$ & $1.16\times 10^5$ \\
  $a_6$ & $-3.38\times 10^5$  \\
  $a_7$ & $6.88\times 10^5$ \\
  $a_8$ & $-9.70 \times 10^5$ \\
  $a_9$ & $9.27 \times 10^5$ \\
  $a_{10}$ & $-5.71\times 10^5$ \\
  $a_{11}$ & $2.05\times 10^5$ \\
  $a_{12}$ & $-3.24\times 10^4$ \\
  \hline
\end{tabular}
\label{tab:01}
\end{table}

\begin{figure}
\centering
\includegraphics[width=0.52\textwidth]{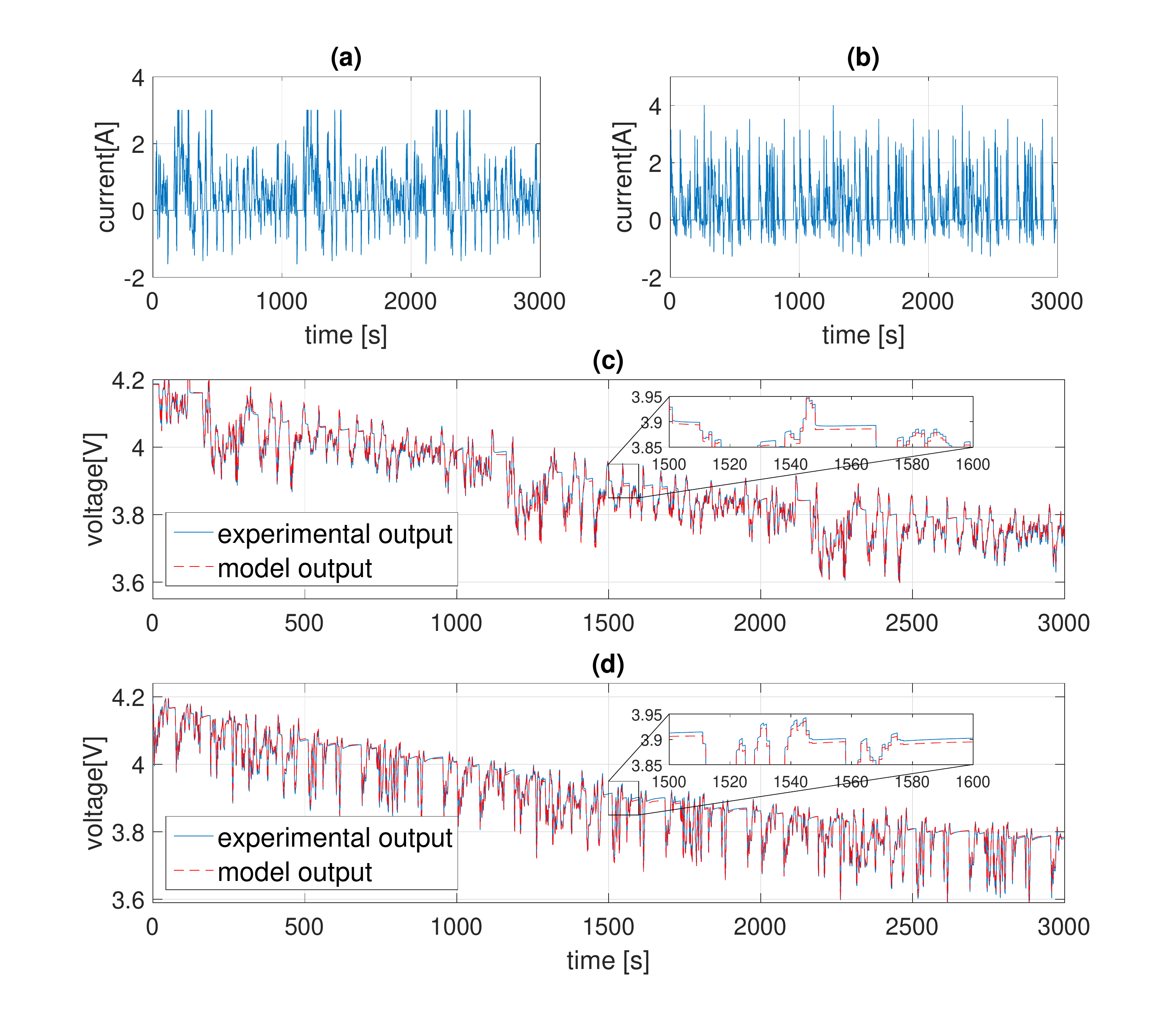}
\caption{Current profiles and model validation results. (a): The FUDS current profile, training dataset; (b): The UDC current profile, validation dataset; (c): Experimental and model outputs under the FUDS load; (d): Experimental and model outputs under the UDC load.}
\label{fig:Figure3}
\end{figure}

Before applying state estimation algorithms to either model (\ref{sys:02}) or model (\ref{sys:2RC_output}), the model parameters need to be identified. We apply a
federal urban driving schedule (FUDS) current profile \cite{testmanual} shown in Fig. \ref{fig:Figure3}(a) to the cell with $\SI{100}{\percent}$ initial SOC and record the applied current and terminal voltage. The parameters are identified using MATLAB\textsuperscript {\circledR}'s System Identification Toolbox and are given in Table \ref{tab:02}.  The voltage measurements and the predicted output by the model under the FUDS load are plotted in Fig. \ref{fig:Figure3}(c). To further validate the model, an ARTEMIS European urban driving cycle (UDC) \cite{Andre2004} shown in Fig. \ref{fig:Figure3}(b) is applied to the same fully charged cell and simulation results under the UDC load using the identified parameters are compared to the experimental voltage measurements, as shown in Fig. \ref{fig:Figure3}(d). The maximal error is $\SI{21.1} {\milli\volt}$ and the RMSE is $\SI{4.78} {\milli\volt}$, which corresponds to $\SI{0.13}{\percent}$ of the mean battery terminal voltage.

\begin{table}
\centering
\caption{The values of model parameters.}
\begin{tabular}{|c|c|}
\hline
 parameter & value \\
  \hline
  $R_1$ & $\SI{2.85 e-2} \ohm$ \\
  $R_2$ & $\SI{4.44 e-2} \ohm$ \\
  $C_1$ & $\SI{4.78 e2} \farad$ \\
  $C_2$ & $\SI{1.83 e4} \farad$ \\
  $R_s$ & $\SI{5.55 e-2} \ohm$ \\
  \hline
\end{tabular}
\label{tab:02}
\end{table}

\subsection{State estimation using various Kalman filters}

To verify that the SOC can be estimated in the presence of unknown sensor biases, we implement several nonlinear filters to model (\ref{sys:2RC_output}) using the experimental measurements of the cell under the UDC load. The biased voltage measurements are obtained by adding a constant to the recorded voltage measurements. The constant is the voltage sensor bias, which is considered unknown at the beginning of the state estimation process. The SOC calculated by Coulomb counting using the accurate lab equipment is regarded as the true SOC \cite{weng2014unified}.

The filtering algorithms implemented here are the first-order EKF, the second-order EKF and the UKF. Since these algorithms are well known and widely used for battery state estimation, the details are not discussed here and we refer the reader to \cite{simon2006optimal} for a comprehensive introduction. The tuning parameters of the filters include the initial state error covariance $P_0$, the process noise covariance $Q_0$ and the measurement noise covariance $R$. For the UKF, there is an extra parameter $\kappa$, which is used to tune the weights of the sigma points \cite{uhlmann2000new}.

\begin{figure}
\centering
\includegraphics[width=0.52\textwidth]{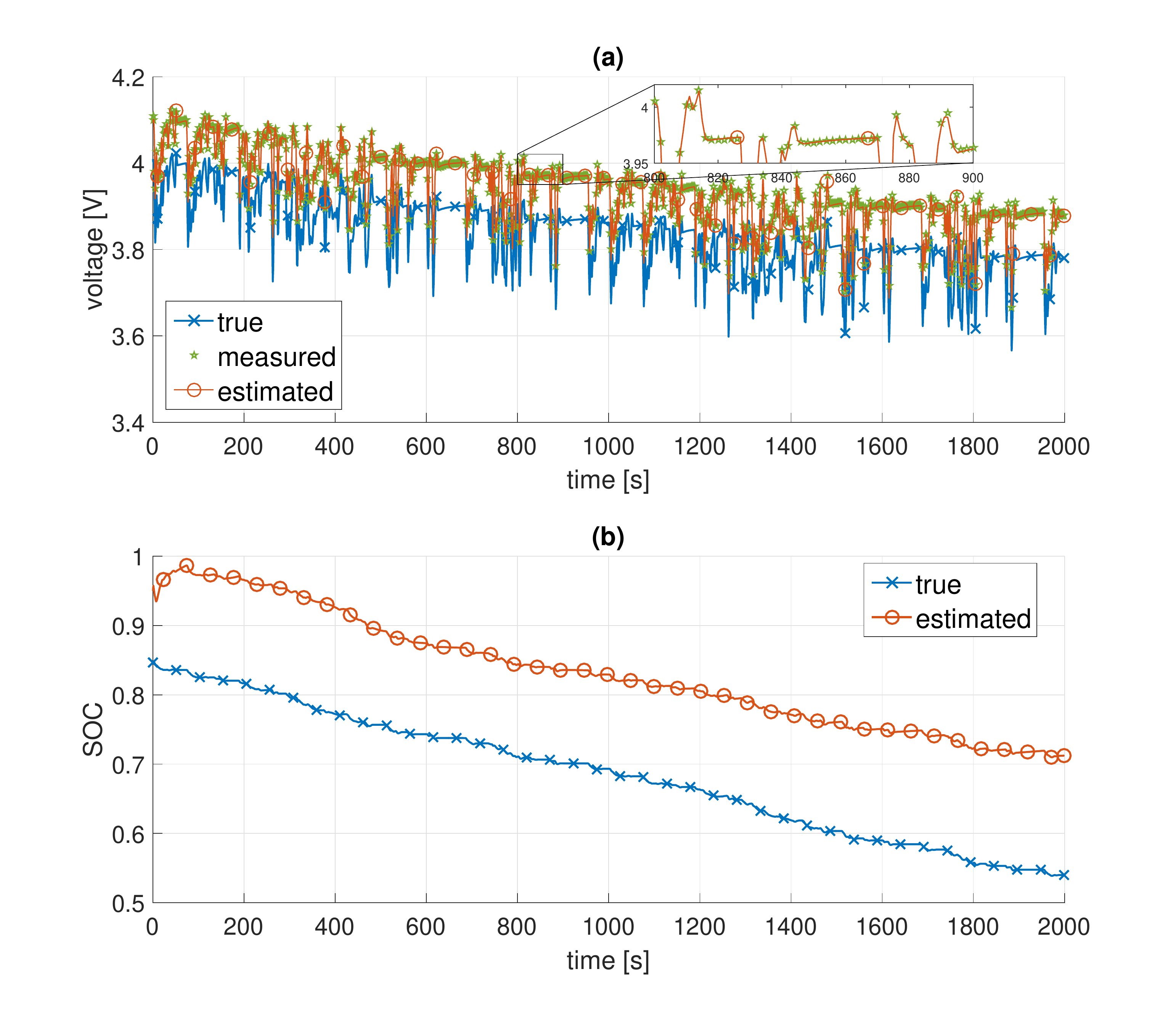}
\caption{State estimation based on model (\ref{sys:02}) using the first-order EKF when there is a bias in the voltage sensor. (a): Evolution of the true battery terminal voltage (blue), the measured voltage (green) and the prediction by the EKF (red); (b): Evolution of the true SOC (blue) and the estimated SOC (red).}
\label{fig:Figure4}
\end{figure}

Consider the case that there is a bias $V_e = \SI{100} {\milli\volt} $ in the measured voltage. 
After one complete driving cycle ($\SI{1000}{\second}$), the true SOC of the cell is about $\SI{85} {\percent}$ and the corresponding OCV is $\SI{4.025}{\volt}$. Suppose the state estimation process starts at this point and due to the presence of the sensor bias, the initial value of the estimation on the SOC is set to $\SI{95} {\percent}$, which means that there is an error of about $\SI{10}{\percent}$ at the beginning. The guessed initial SOC corresponds to an OCV of $\SI{4.123}{\volt}$, which is $\SI{98}{\milli\volt}$ higher than the true value. If the estimation is based on model (\ref{sys:02}), which means that the inaccurate measurement is trusted, it is not surprising that the estimated states are far from their actual values. As shown in Fig. \ref{fig:Figure4}, the SOC estimated by the first-order EKF has considerable errors, due to the biased voltage measurement on which the estimation is based. Note that the estimated voltage in Fig. \ref{fig:Figure4}(a) is calculated from the output equation (\ref{sys:02b}) using the \emph{a posteriori} state estimate. The estimated SOC has a RMSE of $\SI{15.6}{\percent}$ between  $t = \SI{1000}{\second}$ and $ t = \SI{2000}{\second}$.
Using the second-order EKF or the UKF does not improve the estimation accuracy as the sensor bias is still present.

\begin{figure}
\centering
\includegraphics[width=0.52\textwidth]{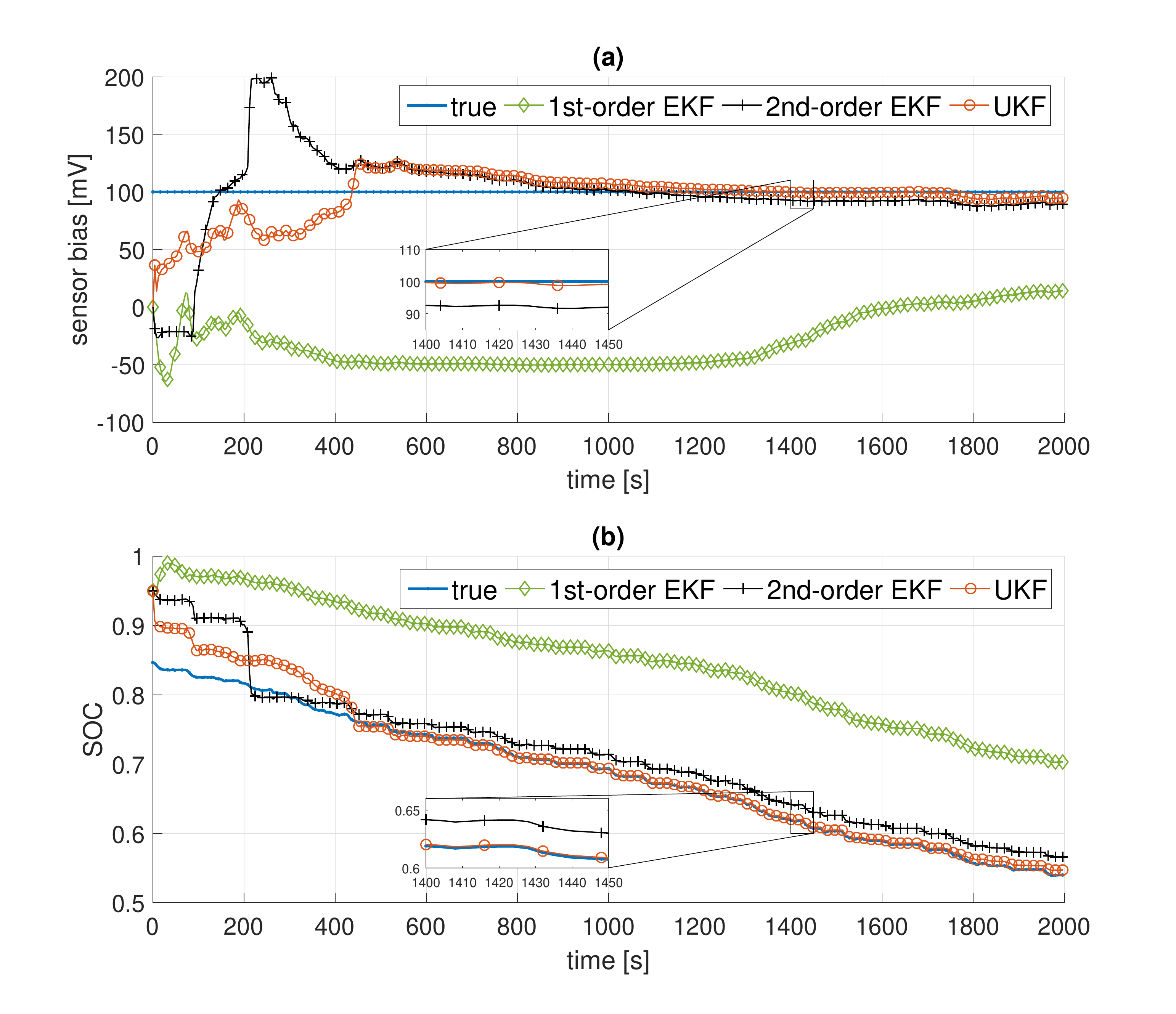}
\caption{State estimation based on model (\ref{sys:2RC_output}). (a): The voltage sensor bias (blue) and the estimated bias by the first-order EKF (green), the second-order EKF (black) and the UKF (red); (b): Evolution of the true SOC (blue) and the estimated SOC by the first-order EKF (green), the second-order EKF (black) and the UKF (red).}
\label{fig:Figure5}
\end{figure}

The analysis in Section \ref{sec:4.2} gives a sufficient condition for model (\ref{sys:2RC_output}) to be locally observable at a point. It can be easily verified that this condition is satisfied everywhere for the polynomial function $V_{OC}$ considered here. Thus it is possible to estimate the SOC and the bias at the same time based on inaccurate measurements.
However, the first-order EKF still performs badly even though the estimation is now based on model (\ref{sys:2RC_output}), as shown in Fig. \ref{fig:Figure5}. The initial error covariance $P_0$ is taken as a diagonal matrix with diagonal entries $\begin{bmatrix} 0.01 & 0.0016 & 0.01 & 0.0625 \end{bmatrix}$, the process noise $Q_0$ is taken as a diagonal matrix with vector $\begin{bmatrix} 10^{-8}  & 10^{-8}  & 10^{-8}  & 10^{-8}  \end{bmatrix}$ 
in the diagonal, the measurement noise standard deviation is taken as $\sigma_V = \SI{6} {\milli\volt}$, which means that $R = 3.6 \times 10^{-5}$, and the time step is set to $\SI{1} {\second}$. The initial value of the estimated bias is set to zero, which means that it is estimated that there is no bias at the beginning. One might attribute the failure of the EKF to poor implementation such as inappropriate initial values for the covariance matrices. However, tuning the EKF does not seem to improve the estimation accuracy much.

The observability analysis of model (\ref{sys:2RC_output}) shows that it is locally observable because of the presence of nonlinearity. It is easy to verify that the observability is lost once the model is linearised. In the first-order EKF, the estimation at every time step is based on a linearisation of the model. Thus it is not surprising that the battery states cannot be estimated accurately.

The estimation accuracy may be improved by using the second-order EKF, which retains the quadratic terms in nonlinear functions. Fig. \ref{fig:Figure5} shows that the second-order EKF with the same filter parameters does have some advantages over the first-order EKF for this problem.

Among the three types of nonlinear Kalman filters, the UKF, which makes use of the unscented transform to account for the nonlinearity, provides the most accurate estimation. With the same filter parameters and $\kappa=4$, the estimated bias soon converges to the true value and its RMSE between $t = \SI{1000}{\second}$ and $ t = \SI{2000}{\second}$ is $\SI{3.41}{\milli\volt}$, which is of the order of the voltage measurement noise standard deviation. The SOC estimation error quickly decreases to less than $\SI{1} {\percent}$ and the RMSE between $t= \SI{1000}{\second}$ and $t= \SI{2000}{\second}$ is $\SI{0.33}{\percent}$. We note that the results are very similar when the voltage sensor bias $V_e$ is negative or zero or when the initial guess of the SOC is smaller than the true value.

Bias in the current sensor has smaller impact on the performances of the nonlinear filters. When there is only a bias $I_e = \SI{-100} {\milli\ampere}$ in the current measurement, the SOC estimation given by the first-order EKF based on model (\ref{sys:02}) has a RMSE of $\SI{2.80}{\percent}$ between $t= \SI{1000}{\second}$ and $t= \SI{2000}{\second}$. This means that the state estimation algorithm is relatively robust against current sensor biases \cite{KleeBarillas2015}. Nevertheless, the estimation accuracy can be further improved if the UKF is applied to model (\ref{sys:2RC_input}), with the RMSE of the estimated SOC in the same time range being reduced to $\SI{0.43}{\percent}$. Furthermore, the current sensor bias $I_e$ can be tracked accurately by the UKF. As a comparison, the second order EKF based on (\ref{sys:2RC_input}) has a similar accuracy and the first order EKF is the least accurate in estimating the SOC and the bias.

The purpose of this section is not to claim that the UKF is always superior to the first order EKF, for if there are no sensor biases, then the first-order EKF based on model (\ref{sys:02}) gives as accurate an estimation as the UKF does. The objective is rather to show that it is important to analyse the observability of a model before choosing an estimation algorithm, with particular relevance to battery SOC estimation. Also, it presents a method to estimate the sensor bias without incorporating any additional sensors into the system.

Note that in the case that there are biases in both the voltage and current sensors, 
{it is probably not possible} to accurately estimate the SOC and the biases at the same time using an augmented model, even when the model appears locally observable. This might be because usually the observability matrix is ill-conditioned, despite of its full rank. 
{Intuitively this makes sense: If we have both a current sensor bias and a voltage sensor bias, there is ambiguity regarding whether an error in estimated model output compared to measured output is caused by the current sensor error or the voltage sensor error.}

\section{Conclusion} \label{sec:conclusion}
The importance of observability analysis of nonlinear ECMs for battery state estimation has not received much attention. This brief analyses the nonlinear observability of a widely used second-order RC model for lithium-ion batteries. Using the ECM as an example, it is pointed out that the local observability of a nonlinear dynamic system at a certain point is not the same as the observability of the system linearised around the point. We then present a method to estimate the battery SOC when there are unknown biases in the sensors. The estimation is based on an augmented model which incorporates the sensor biases into the state vector. The observability of the augmented model is shown to be dependent on the model nonlinearity. Experimental validation shows that it is vital to take the nonlinearity of the model into account in the estimation algorithm, highlighting the importance of observability analysis for state estimation. We also show that accurate voltage measurement is critical for SOC estimation in a BMS using the original model, whereas current measurement accuracy has a smaller impact as the Kalman filters are relatively robust against current sensor bias (due to their knowledge of the OCV-SOC `ground truth' and ECM parameters).
{If the model parameters and OCV curve are prone to change over time, this makes the problem significantly more challenging, but is a topic for future investigation.}

\section*{Acknowledgements}
The authors would like to thank Christoph Birkl for his help with experiments.

\end{document}